\documentclass[epj]{svjour}
\usepackage{graphicx}
\usepackage[latin1]{inputenc}
\usepackage{amsmath}

\begin{document}
\newcommand{\Zmoy}{\langle Z \rangle}
\newcommand{\Zbound}{Z_{\rm bound}}
\newcommand{\rd}{\mathrm{d}}

\title {Fragment Charge Correlations and Spinodal Decomposition in Finite 
Nuclear Systems}

\author{
G.~T\u{a}b\u{a}caru\inst{1,2}\thanks{\emph{Present address:} Cyclotron Institute, 
Texas A\&M University, College station, Texas 77845, USA}
\and B.~Borderie\inst{1}\mail{borderie@ipno.in2p3.fr}
\and P.~Désesquelles\inst{1}
\and M.~P\^arlog\inst{1,2}
\and M.F.~Rivet\inst{1}
\and R.~Bougault\inst{3}
\and B.~Bouriquet\inst{4}
\and A.M.~Buta\inst{3}
\and E.~Galichet\inst{1,5}
\and B.~Guiot\inst{4}
\and P.~Lautesse\inst{6}
\and N.~Le~Neindre\inst{4}\thanks{\emph{Permanent address:} Institut 
de Physique Nucl\'eaire, IN2P3-CNRS, F-91406 Orsay cedex, France.}
\and L.~Manduci\inst{3}
\and E.~Rosato\inst{7}
\and B.~Tamain\inst{3}
\and M.~Vigilante\inst{7}
\and J.P.~Wieleczko\inst{4}
}
\institute{
Institut de Physique Nucl\'eaire, IN2P3-CNRS, F-91406 Orsay
 cedex, France
\and National Institute for Physics and Nuclear Engineering,
RO-76900 Bucharest-M\u{a}gurele, Romania.
\and LPC, IN2P3-CNRS, ENSICAEN et Universit\'e, F-14050 Caen cedex, France.
\and GANIL, CEA et IN2P3-CNRS, B.P.~5027, F-14076 Caen cedex, France.
\and Conservatoire National des Arts et M\'etiers, F-75141 Paris cedex 03,
France.
\and Institut de Physique Nucl\'eaire, IN2P3-CNRS et Universit\'e,
F-69622 Villeurbanne cedex, France.
\and Dipartimento di Scienze Fisiche e Sezione INFN, Universit\'a
di Napoli ``Federico II'', I80126 Napoli, Italy.
}
\date{\today}
\abstract{
Enhanced production of events with almost equal-sized fragments is
experimentally revealed by charge correlations in the multifragmentation of 
a finite nuclear system selected in $^{129}$Xe central collisions on 
$^{nat}$Sn. The evolution of their weight with the incident energy: 32, 39, 
45, 50 AMeV, is measured. Dynamical stochastic mean field simulations 
performed at 32 AMeV, in which spinodal instabilities are responsible for 
multifragmentation, exhibit a similar enhancement of this kind of events. 
The above experimental observation evidences the spinodal decomposition of
hot finite
nuclear matter as the origin of multifragmentation in the Fermi energy regime.
}
\PACS{
 25.70.Pq \and 24.60.Ky }
 
\maketitle
\section{Introduction\label{intro}}
When enough energy is brought into a nucleus, it breaks up into smaller
pieces: this is called multifragmentation~\cite{MoWo93,Gro90}. The knowledge 
of this process, experimentally observed for many years, was recently strongly 
improved with the analysis of experiments performed with powerful detection
arrays~\cite{MDA00,Hau00,Elli00,Xu00,I29-Fra01,Bot01,MDA02,Radu02,Kle02,Ell02}.
As a theoretical framework for describing  this phenomenon, the analysis 
of the bulk dynamics of nuclear matter, under various conditions of density 
and internal energy, has found arguments that multifragmentation occurs
when nuclear matter has expanded through the spinodal region of negative 
compressibility~\cite{Ber83}. In this area of mechanical instabilities, 
covering a large part of a liquid-gas type coexistence domain, the 
irreversible growth of local density fluctuations leads the system to the 
separation into two phases: the spinodal decomposition. For the liquid part 
primary fragments of nearly equal sizes should be favoured, in relation to 
the wavelengths of the most unstable modes~\cite{Ayi95}. Effects like beating 
of different modes, coalescence of nascent fragments, secondary decay of the 
excited fragments and, above all, finite size effects are expected to deeply 
blur this simple picture~\cite{Jac96,Colo97}. However, evidence for an
enhancement of events with nearly equal-sized fragments was for the first time
observed in 32 AMeV $^{129}$Xe + $^{nat}$Sn central collisions leading to a 
fused system which undergoes multifragmentation, by using a charge
correlation method~\cite{I31-Bor01}. 
By means of this model independent method, recently improved~\cite{Des02}, 
we are investigating here the evolution of the equal-sized fragment partitions
with the increase of the incident energy up to 50 AMeV, in selected samples 
of experimental fused events concerning the same system. In the same framework,
for collisions at the lowest energy, the predictions of the 3D stochastic 
mean-field simulations are successfully compared to the experimental results;
these simulations take into account the dynamics of the most unstable modes 
in the spinodal region.
\par
The paper is organized as follows. In section 2 we firstly present the 
experimental set-up, including the detector array specifications and 
operating conditions during experiment. Secondly, we recall the criteria 
allowing to select the experimental events corresponding to  fused systems 
in central collisions. The general features of the related  experimental 
data are evidenced in section 3. In section 4, the charge correlation method 
with some improvements and developments is described; results of stochastic 
mean field simulations are used to
exemplify and to compare the potentiality of the up-dated version  
to the original method. The experimental results, obtained at different
incident energies, are then shown; they are discussed and interpreted in 
section 5. Conclusions of this study are drawn in section 6.

\section{Experimental selection of a finite piece of nuclear matter\label{sect2}}
Heavy ion collisions at relative velocities comparable to those of
the nucleons in the nucleus may provide test pieces of nuclear matter at
moderate temperatures which are good candidates to undergo a nuclear 
liquid-gas type phase transition. Indeed, ``fused'' systems reaching sizes 
up to a few hundreds of nucleons may be obtained in the laboratory,
by colliding accelerated heavy projectiles with heavy targets. 
By means of $4\pi$ detection arrays of high granularity,
it is possible to completely study their disassembly.
Such events are expected to appear with high probability for central
collisions, but one major problem is to select them among 
reactions with a dominant binary character~\cite{I28-Fra01}.
\subsection{Experimental procedure\label{experiment}}
The $^{129}$Xe + $^{nat}$Sn system was studied with the $4\pi$ multidetector
INDRA, operating at the GANIL accelerator. A thin target of natural tin 
(350 $\mu$g/cm$^{2}$) was bombarded by $^{129}$Xe projectiles at five 
incident energies: 25, 32, 39, 45 and 50 AMeV. Low intensity beams 
($\sim$ 3 $\times$ 10$^7$ pps) were used to keep the random coincidence rate 
below 10$^{-4}$. The low target thickness allows slow fragments to escape 
the target. A trigger based on the multiplicity was chosen, requiring 
at least four modules firing. In the off-line analysis, events having a 
multiplicity of correctly identified charged particles inferior to the 
experimental trigger condition were rejected for reasons of coherency. 
The most peripheral collisions are thus eliminated.

INDRA, which is described in detail in~\cite{I3-Pou95,I5-Pou96}, can be 
viewed as an ensemble of 336 telescopes covering about 90\% of the $4\pi$ 
solid angle. The detection cells are distributed amongst 17 rings centred 
on the beam axis. Low-energy identification thresholds (from 
$\approx 0.7 A$MeV for $Z = 3$ to $\approx 1.7 A$MeV for $Z = 50$) and large 
energy ranges were obtained through the design of three layer telescopes, 
composed of an axial-field ionization chamber operated at 30 mbar of
$C_{3}F_{8}$, a 300 $\mu$m silicon detector and a CsI(Tl) scintillator, 
thick enough to stop all emitted particles, coupled to a phototube. Such 
a telescope can detect and identify from protons between 1 and 200 MeV 
to uranium ions of 4 GeV. Past 45$^\circ$, where fast projectile-like 
fragments are no longer expected, the telescopes comprise only two stages, 
the ionization chamber operated at 20 mbar and the scintillator. Finally 
the very forward angles (2 - 3$^\circ$) are occupied by NE102 - NE115 
phoswiches. 

A charge resolution of one unit was obtained for the whole range of atomic 
number of fragments identified through $\Delta E - E$ method in the 
Si - CsI(Tl) couple. For CsI(Tl) scintillators a better understanding of
the light response was obtained and the contribution of $\delta - rays$
generated by the incoming heavy ion was taken into account \cite{I33-Par02}. A
direct consequence was the identification of fragment with a resolution of
one charge unit up to $Z = 20$, and a few charge units  for the heaviest 
fragments, in ionization chamber - CsI(Tl) telescopes \cite{I34-Par02}. 
The exact identification of fragments up to at least $Z = 20$ reveals 
essential for the charge correlation studies which are the aim of this paper.

\subsection{Selection of single source events\label{subsect22}}

A two step procedure has been used to isolate fused systems.
The first step was to keep the events for which a quasi-complete detection of 
the reaction products has been achieved. Significant fractions: $\geq 77 \%$ 
of the charge of the system, $Z_{sys} = Z_{proj} + Z_{targ}$, and $\geq 75 \%$
of the beam momentum, $P_{proj}$, in the exit channel were required to be 
measured for every event. INDRA does not permit isotopic identification for 
fragments and does not detect the neutrons. For this reason, the momentum 
used here is calculated from the product of atomic number $Z$ and velocity 
component in the beam direction $v_{z}$: $P_{tot} = \Sigma Zv_{z}$ and 
normalized to the incident (projectile) momentum: $Z_{proj}v_{proj}$.
In the second step we used the flow angle ($\Theta_{flow}$) 
selection~\cite{Cug83,Lec94,I28-Fra01}. This global variable is defined as the
angle between the beam axis and the preferred direction of emitted matter
in each event. It is determined by the energy tensor calculated  from 
fragment ($Z \geq 5$) momenta in the reaction centre of mass. Fused events 
have no memory of the entrance channel and should be isotropic while binary 
dissipative collisions are focused at small $\Theta_{flow}$. Thus, by 
selecting only large flow angles, fused events can be well isolated.
In~\cite{I28-Fra01} the minimum flow angle chosen for 32 AMeV collisions was
70$^\circ$. It was 60$^\circ$ for 50 AMeV collisions~\cite{I9-Mar97}.
In this paper we chose 60$^\circ$ for all energies, to get enough
statistics (at least 30000 events) without degrading the properties of a
compact fused source. 

The present selection corresponds to measured cross sections decreasing 
from 56 to 19~mb when the incident energy goes from 25 to 50 AMeV 
(see table~\ref{table1}). 
By taking into account detection efficiency and biases due to the selection 
(quasi-complete events and flow angle selection) the total cross section for 
the formation of compact fused systems is estimated to decrease from 250 to
85~mb between 25 and 50 AMeV~\cite{I39-Hud02}. 

\section{Global properties of selected fused events\label{centcol}}
\begin{table*}[htb]
\caption{Exit channel average values of the total and fragment ($Z \geq 5$)
multiplicity, total charge emitted in fragments and fragment atomic number, 
in events from $^{129}$Xe + $^{nat}$Sn central collisions, as functions of 
the entrance channel characteristics: incident and available  energy. The 
numbers in parentheses are the standard deviations of the distributions. The
measured cross sections are given in the last column.}
\begin{center}
\begin{tabular}{lcccccc}
\hline
   $E$ & $E_{cm}/A_{sys}$  & $\langle M_{tot}\rangle$ & $\langle M_f\rangle$ &
   $\langle \Zbound\rangle$ & $\langle Z_f\rangle$ & $\sigma_{meas.}$ \\
   AMeV & AMeV & & & & &  mb\\
\hline
 25 &  6.24 &  18.9 (2.7) & 3.56 (1.04) & 67.3 (6.8) & 17.6 & 56. $\pm$ 8. \\
 32 &  7.99 &  23.8 (3.0) & 4.13 (1.17) & 57.6 (7.0) & 13.4 & 26. $\pm$ 4. \\
 39 &  9.73 &  28.5 (3.1) & 4.41 (1.20) & 49.6 (7.4) & 11.0 & 18. $\pm$ 3. \\
 45 & 11.23 &  31.5 (3.3) & 4.42 (1.19) & 44.0 (7.5) &  9.7 & 19. $\pm$ 5. \\
 50 & 12.48 &  34.2 (3.3) & 4.31 (1.18) & 39.7 (7.6) &  9.0 & 19. $\pm$ 3. \\
\hline
\end{tabular} \\
\end{center}
\label{table1}
\end{table*}
Before presenting and discussing charge correlations, which deal with 
information inside events, it is useful to have an overview of the related 
inclusive properties of the selected events, like the multiplicity and charge 
distributions. Entrance channel conditions for the Xe+Sn system of total mass 
$A_{sys} = 248$ and charge $Z_{sys} = 104$ are summarized in 
Table~\ref{table1}. 

\subsection{Multiplicity and bound charge\label{MultZb}}

\begin{figure*}[htb]
\centering
\includegraphics{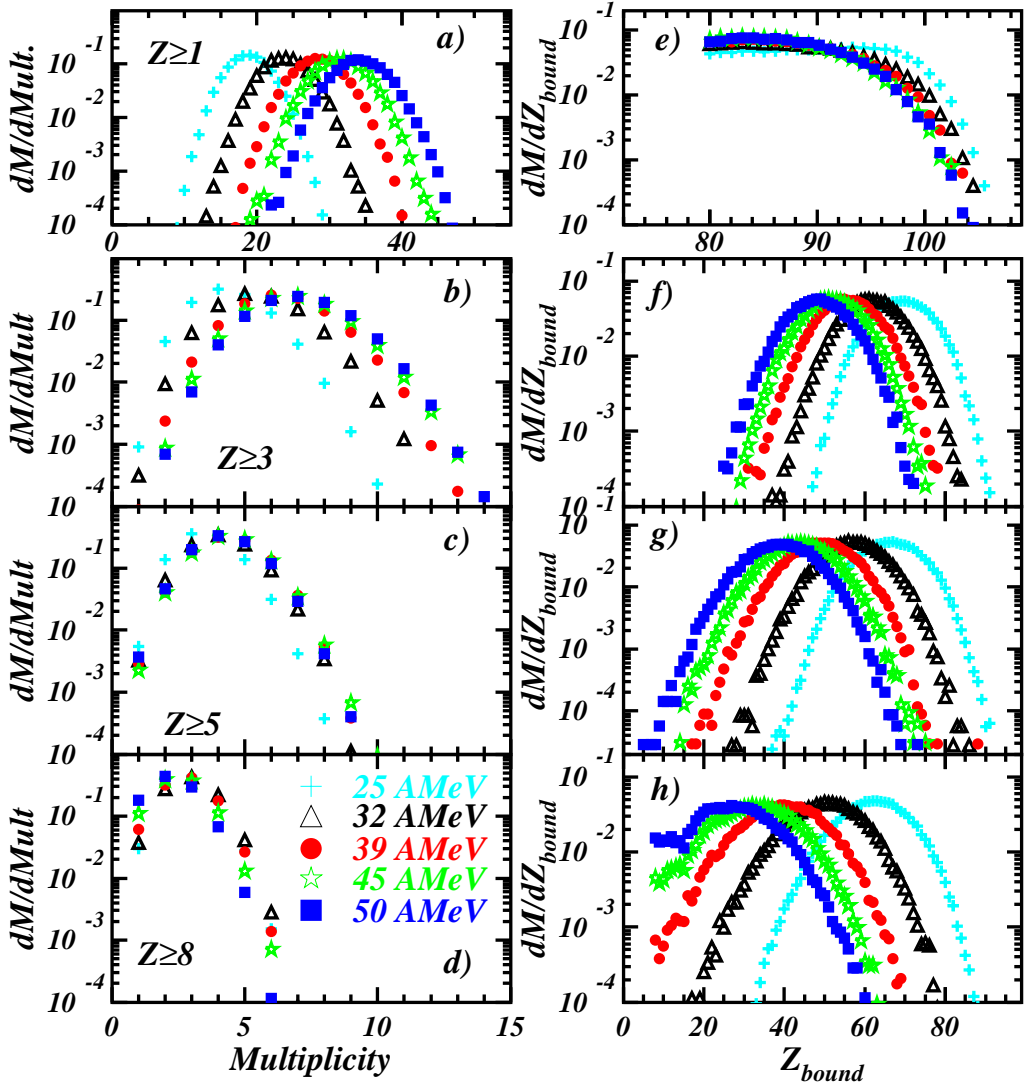}
\caption{Experimental multiplicity (left) and total charge (right)
distributions for the selected events from central 25, 32, 39, 45 and 50
AMeV $^{129}$Xe + $^{nat}$Sn collisions, with different lower thresholds
$Z_{min}$ for the reaction products taken into account. From top to bottom:
$Z_{min} = $1, 3, 5, 8. Note that panel e) thus displays the total detected 
charge.
}
\label{mulzb}
\end{figure*}
Fig.~\ref{mulzb} shows multiplicity and total charge
($\Zbound$ = \linebreak $\sum_{Z \geq Z_{min}} Z$) distributions normalized
to the number of events, at the five incident energies. 
The total multiplicity distributions, when all the charged reaction 
products are considered, $Z \geq 1$, are shown in fig.~\ref{mulzb}a. 
They all display a gaussian shape; the most probable multiplicity values
and the standard deviations increase with the available energy, as shown in
Table~\ref{table1}. At 50 AMeV the average multiplicity reaches one third of the total
charge.  The distributions of the sum of the charges of all the reaction 
products, plotted in fig.~\ref{mulzb}e), for different
energies, are practically identical, nicely testifying about the negligible
variation of the detection efficiency of INDRA when the incident energy
significantly varies. The common limitation at 80 charge units is demanded by
software as a criterion for the completeness of event detection. These two
plots, considered together, speak about the increasing number of reaction
products from central collisions with the incident energy.

The evolution of the reaction product multiplicity distribution, at all
energies, when another lower charge limit is imposed, is presented afterwards
in the left column of the same figure. As compared to fig.~\ref{mulzb}a 
($Z \geq 1$), the splitting of the energy dependent distributions is 
diminishing if $Z \geq 3$ (fig.~\ref{mulzb}b); the curves become completely 
superimposable, except at 25 AMeV,  for $Z \geq 5$ (fig.~\ref{mulzb}c) and 
split again, but in the opposite sense, if $Z \geq 8$ (fig.~\ref{mulzb}d). 
Deexcitation simulations for primary fragments having the same N/Z ratio 
as the total system, performed by means of the statistical code 
SIMON~\cite{T18ADN98} at 3 AMeV excitation energies, have indicated small
evaporation rates of Li and Be isotopes, but negligible rates of heavier 
nuclei~\cite{T27Tab00}. This average value of 3 AMeV for the fragment 
excitation energy was deduced from fragment-particle correlations for the 
same system between 32 and 50 AMeV incident energy~\cite{I11-Mar98,I39-Hud02}.
Moreover the daughter nucleus atomic number is, on average, one charge unit
smaller than that of the parent fragment, with a standard deviation of about
one unit. It may be thus inferred that the distribution of the number of
primary fragments with $Z \geq 5$ is not dramatically modified by the 
secondary decay. More precisely, as
indicated in Table~\ref{table1}, the average number of fragments with 
$Z \geq 5$ reaches a maximum for an incident energy as low as 39 AMeV and
starts decreasing above 45 AMeV.

 If the lower limit is set to $Z_{min}$=3, for comparison with other works, 
the maximum value, 7.1 fragments, is reached at 50 AMeV. The same maximum 
value is obtained for the asymmetric Ni+Au collisions measured with
INDRA~\cite{I37-Bel02}. These results can be compared with a series of
measurements of fragment multiplicities published in~\cite{Sis01}. Using for
the present data a similar selection method would  increase the INDRA value by
$\sim$10\%. These INDRA data, combined with  those related to Ni+Ni and Kr+Nb
of~\cite{Sis01} show that the maximum  average number of fragments produced in
central collisions  (multifragmentation) is proportional to the total system
mass A$_{sys}$ up  to at least A$_{sys} \sim$250; there is no indication of a
saturation,  as inferred from the Kr+Au data of~\cite{Wil97} and from
percolation  calculations. The maximum number of fragments for A$_{sys}
\sim$250 is  moreover reached for a c.m. available energy of $\sim$ 12 AMeV  in
the INDRA measurements, much lower than those quoted in~\cite{Sis01}. 

Pictures f), g), h) in the right column of fig.~\ref{mulzb} show 
the distributions of the sum of the charges of the reaction products 
having atomic numbers $Z \geq $ 3, 5, and 8, respectively. 
The independence of the energy, seen in fig.~\ref{mulzb}e), is progressively
removed from top to bottom and the populated domain shifts
towards lower values of the total charge of the reaction products taken into
account. For $Z \geq 5$, for example, the shapes of the distribution are
very alike: gaussians which are slightly broadening with increasing energy,
but centred at lower and lower values of the sum $\Zbound$ of the fragment
charges. For $Z \geq 8$ and the largest incident energies (45 and 50 AMeV)
shoulders are observed at low $\Zbound$, which indicates the onset of events
with a high degree of fragmentation:
those events have only one fragment with a charge in the range 8-14. 

\par The evolution of the exit channel average measured quantities with the
available energy  is also synthesized in Table~\ref{table1}.  These average
values concern: the total multiplicity, $M_{tot}$, the  fragment ($Z \geq 5$)
multiplicity, $M_f$, the total charge emitted in  fragments, $\Zbound$, and the
fragment atomic number, $Z_f$. The growth  of the average total multiplicity,
when the incident energy increases, is accounted by the growth of the
light nuclei ($Z \leq 4$)  average multiplicities, while the average fragment
multiplicity barely changes. The ratio of the bound charge to the total
charge of the system  goes down with increasing excitation energy. The size of
the produced  fragments is diminishing too, as a proof that the collisions are
becoming  more violent. At a given incident energy,
 however, $\Zbound$ increases roughly linearly as a function of fragment
(Z$\geq 5$)  multiplicity with a slope varying from 3 to 4.5 when the energy
increases from 25 to 50 AMeV. This again indicates that at higher energy the
system is highly fragmented.

\subsection{Charge distributions\label{subsect33}}
\begin{figure}[htb]
\includegraphics[scale=0.75]{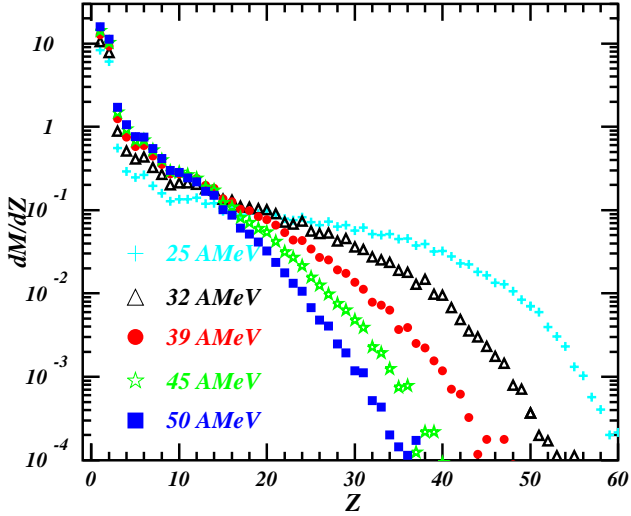}
\caption{Experimental differential charge multiplicity distributions
for the selected events formed in central 25, 32, 39, 45 and 50
AMeV $^{129}$Xe + $^{nat}$Sn collisions.}
\label{fig2}
\end{figure}
The charge distributions, normalized to the number of single source events,
can be compared all together in fig.~\ref{fig2}. When the available energy
increases, the emission of small fragments ($Z < 10$) increases while that
of fragments with $Z > 15$ decreases producing steeper and steeper 
distributions;
quite equivalent rates of production for fragments with
 $10 \leq Z \leq 15$ are observed between 32 and 50 AMeV.
 The domain of the heaviest fragments $Z \approx 50$, populated at the 
lowest energy, gradually vanishes at higher energies. This behaviour is 
quite remarkable, particularly seeing  that the number of fragments is 
the same at all energies (fig.~\ref{mulzb}c).
Roughly, one observes the following trend: the curves are gradually
passing from a regime with two slopes, at the lowest incident energy,
towards a regime with one slope, at the highest incident energy.
The average measured charge of the fragments can be found in 
Table~\ref{table1}.

\section{Charge correlation functions and enhanced production of nearly
equal-sized fragment events\label{sect4}}
A few years ago a new method called higher order charge
correlations~\cite{Mor96} was proposed to enlighten any extra production of
events with specific fragment partitions. The high sensitivity of the method
makes it particularly appropriate to look for small numbers of events as those
expected to have kept a memory of spinodal decomposition properties.
Thus, such a charge correlation method allows to examine model independent
signatures that would indicate a preferred decay into a number of equal-sized
fragments in events from experimental data or from simulations. In this
section and the following only events with 3 to 6 fragments (Z $\geq$ 5) at
32, 39, 45 and 50 AMeV will be considered. They represent in all cases 
90-92 \% of the selection previously considered. At these four energies, the
detection efficiency is independent of the fragment charge, in the selected
event samples. This is not so well verified at 25 AMeV.

\subsection{Methods\label{subsect41}}
\par
The classical two fragment charge correlation method considers the  coincidence
yield $Y(Z_1,Z_2)$ of two fragments of atomic numbers $Z_{1,2}$,  in the events
of multiplicity $M_f$ of a sample. A background yield  $Y'(Z_1,Z_2)$ is
constructed by mixing, at random, fragments from different  coincidence events
selected by the same cut  on $M_f$. The two particle  correlation function is
given by the ratio of these yields. When searching for enhanced production of
events which break into equal-sized fragments, the higher order correlation
method appears much more sensitive. All fragments of one event with fragment
multiplicity $M_f$ = $M = \sum_Z n_Z$, where $n_Z$ is the number of fragments
with charge $Z$ in the partition,  are taken into account. By means of the
normalized first order:
\begin{equation}
	\Zmoy = \frac{1}{M} \sum_Z n_Z Z
\label{equ1}
\end{equation}
and second order:
\begin{equation}
        \sigma_Z^2 = \frac{1}{M} \sum_Z n_Z (Z - \Zmoy)^2
\label{equ2}
\end{equation}
moments of the fragment charge distribution in the event, 
one may define the higher order charge correlation function:
\begin{equation}
\left. 1+R(\sigma_Z, \Zmoy)=\frac{Y(\sigma_Z, \Zmoy)}{Y'(\sigma_Z, \Zmoy)} 
\right| _{M}
\label{equ3}
\end{equation}

Here, the numerator $Y(\sigma_Z, \Zmoy)$ is the yield of events with given
$\Zmoy$  and $\sigma_Z$ values. Because the measurement of the charge 
belonging to a given event is not subject to statistical fluctuations, we 
use here expression  (\ref{equ2}) rather than the ``nonbiased estimator'' 
of the variance,  $\frac{1}{M-1} \sum_Z n_Z (Z - \Zmoy)^2$,  as proposed in
~\cite{Mor96} and used in our previous paper~\cite{I31-Bor01}. Note that this
choice has no  qualitative influence on the forthcoming conclusions. The
denominator $Y'(\sigma_Z, \Zmoy)$, which represents the uncorrelated yield
of pseudo-events, was built in~\cite{Mor96}, as for classical correlation  
methods, by taking fragments at random in different events of the selected  
sample of a certain fragment multiplicity. This Monte-Carlo generation of 
the denominator $Y'(\sigma_Z, \Zmoy)$ can be replaced by a fast algebraic  
calculation which is equivalent to the sampling of an infinite number of 
pseudo-events~\cite{Des02}. Its contribution to the statistical error of 
the correlation function is thus eliminated.  However, owing to the way the
denominator was constructed, only the  fragment charge distribution 
$\rd M/ \rd Z$ of the parent sample is reproduced  but the 
constraints imposed 
by charge conservation are not taken into account. This has, in particular, 
a strong effect on the charge bound in fragments 
$\rd M/ \rd \Zbound$ 
distribution. This fact makes the denominator yield distributions as
a function of $\Zmoy$ wider and flatter than those of the
numerator~\cite{T27Tab00}.
Consequently, even in the absence of a physical correlation signals, the 
ratio (\ref{equ3}) is not a constant equal to one. The correlations induced 
by the finite size of the system (charge conservation) distorts the amplitude, 
or even may cancel other less trivial correlations. 
Therefore, a new 
method for the evaluation of the denominator~\cite{Des02}, based on 
the ``intrinsic probability'' of emission of a given charge, was adopted.
It minimizes these effects and replicates all features of the partitions 
of the numerator, except those (of interest) due to other reasons 
than charge conservation.

\par
The goal of the method is to take into account in a combinatorial way 
the trivial correlations due to charge conservation.
If there is no correlation 
between the charges, each charge can be fully described by an
emission probability (referred to as intrinsic probability and noted 
$^{\rm intr}P_Z$). This fact was largely demonstrated for the bulk of 
multifragment emission around the Fermi energy~\cite{Mor97}. Without
charge conservation constraint, the intrinsic probabilities would be equal 
to the emission probability. Charge conservation makes the emission 
probability of high charge fragments smaller than their intrinsic probability, 
whereas it is the contrary for small charges. The probability to observe 
a given partition (${\bf n}: (n_1,\ldots ,n_{Z_{\rm tot}})$), at a given total
multiplicity, $M_{tot} = m = \sum_Z n_Z$, is obtained by the multinomial 
formula. If the total charge is fixed ($Z_{\rm tot}=\sum_Z Z\,n_Z$), 
the partition probabilities are given by:
\begin{equation}
\label{Eq P(n)}
P({\bf n}|m) = \alpha \ m! \,
\prod_Z \frac{^{\rm intr}P_Z^{n_Z}} {n_Z!} \
\delta_{Z_{\rm tot},\sum_Z Z\,n_Z}\ ,
\end{equation}
\noindent
where $\alpha$ is the normalization constant  (so that $\sum_n
P({\bf n}|m$) = 1) and $\delta$ is the Kronecker symbol. 
For the reason previously mentioned, no Z-dependent efficiency term is
needed in this formula.
These partition probabilities contain all information relative  to the
charges and their correlations. For example, the denominator $Y'(\sigma_Z,
\Zmoy)$ is obtained by summing the probabilities of the partitions with given
mean charge and standard deviation. Of course, the intrinsic probabilities are
not direct experimental observables, they have to be evaluated by inversion of
Eq. (\ref{Eq P(n)}). However, {\it this  inversion is possible only if the 
physical correlations} (not due to charge conservation) {\it are weak}. If the
data sample contains only trivial correlations, then the higher order charge
correlation  function (or any other correlation function) is everywhere equal
to 1.  Local positive (negative) physical correlations will appear as peaks 
(holes). In the case of experimental events, it has to be noted that  a set of
intrinsic probabilities exists only if the selected sample comprises
essentially single source events having reached thermal equilibrium  (the 
source has to be described by a unique set of intrinsic  probabilities).
Otherwise the results would be equal to the convolution of  the partition
probabilities corresponding to the different source sizes,  each of them being
described by a different set of intrinsic probabilities. Hence, the convergence
of the inversion procedure (Eq. \ref{Eq P(n)}) is a  strong indication that these
two conditions are fulfilled by the sample under study~\cite{Des02}. The data
samples correspond to multifragmenting  single sources, supplemented by
pre-equilibrium and secondary decay particles and light fragments. Hence, only
fragments with charge greater than or equal to 5 have been retained for the
analysis of the multifragmenting source. This feature has to be taken into
account in the calculation of the denominator. The fragment partition (${\bf
N}:(n_5,\ldots,n_{Z_{\rm tot}})$)  probability can be calculated as the sum
over the complete partitions  ${\bf n}$ which include ${\bf N}$. Noting $M$ the
fragment multiplicity, it can be straightforwardly shown from Eq.~(\ref{Eq
P(n)}) that the fragment partition probability reads:

\begin{equation}
\label{Eq P(N)}
P({\bf N}) = f(M,Z_{\rm bound})\,P'({\bf N})\ ,
\end{equation}
with
\noindent \begin{equation}
\begin{array}{l}
\label{Eq f(M,Zbound),P'(N)}
\begin{split}
f(M,Z_{\rm bound})=\alpha \sum_{{\bf n}:(n_1,\ldots,n_4)} 
{m \choose M}
(m-M)!\ \\ 
\times \prod_{Z=1}^4 \frac{^{\rm intr}P_Z^{n_Z}} {n_Z!}  \
\delta_{Z_{\rm tot}-Z_{\rm bound},\sum_{Z=1}^4 Z\,n_Z}
\end{split} \\
P'({\bf N}) = M!\ \prod_{Z=5}^{Z_{\rm tot}} 
\frac{^{\rm intr}P_Z^{n_Z}} {n_Z!}\ ,
\end{array}
\end{equation}

where ${m \choose M}$ are the binomial coefficients. The value of
$Z_{\rm tot}$ was fixed at the total charge of the system,  $Z_{\rm
projectile}$ + $Z_{\rm target}$; it has been noticed that, whereas
the total
charge conservation has to be explicitely included in the calculation, the
results are only modified when $Z_{\rm tot}$ is lower than 85. This new 
method to build the denominator will be denoted as the Intrinsic Probability 
Method (IPM) in what follows.
However,
the explicit calculation of the intrinsic probabilities may not be the only
method for building a denominator including only the correlations induced by
charge conservation (another procedure is proposed in~\cite{I41-Cha02}, see
Appendix for a comparison with the IPM method).
In any case, by definition, all valid methods lead to a partition 
probability described by Eq. (\ref{Eq P(n)}).

\par A comparison of the results obtained in the framework of
the higher order correlation method, with the analytical denominator, and with
IPM, will be  drawn in the following for an event sample resulting from a
simulation of 32 AMeV $^{129}$Xe + $^{119}$Sn central collisions.

\subsection{Stochastic mean-field and spinodal
decomposition\label{subsect42}}

\begin{figure*}[htb] 
\centering
\includegraphics[scale=0.8]{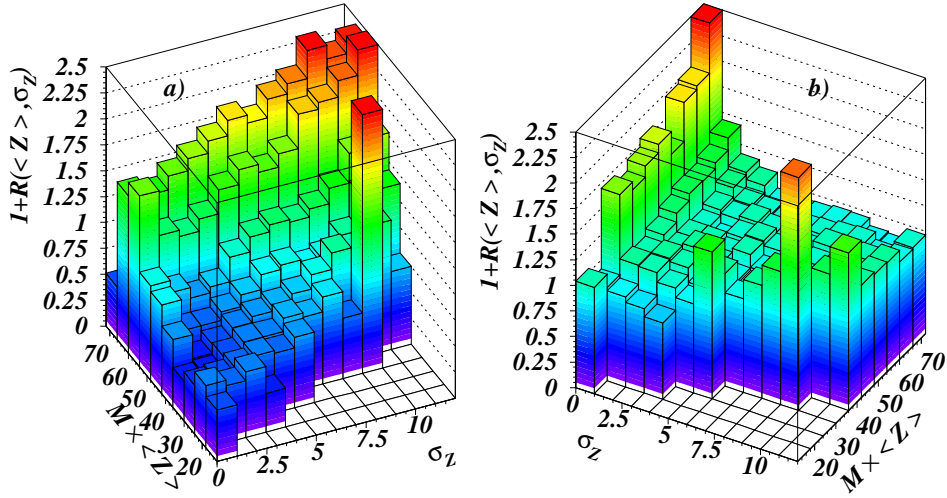}
\caption{Correlation functions for events with $M_f$ = 3 to 6, simulated with 
the Brownian One Body model for 32 AMeV $^{129}$Xe + $^{nat}$Sn collisions.  
a) with an analytical denominator provided by pseudo-events; b) with a 
denominator calculated with the IPM. The orientations
of a) and b) are different for a better visualisation of the landscapes}
\label{fig3}
\end{figure*}
Dynamical stochastic mean-field simulations have been proposed for a long time
to describe processes involving instabilities like those leading to
spinodal decomposition~\cite{Ran90,Cho91,Bur91}.
In this approach, spinodal decomposition of hot and dilute finite nuclear
systems can be mimicked through the Brownian One-Body (BOB) dynamics
~\cite{Cho94,Gua97,I29-Fra01}, which consists in employing a Brownian force 
in the kinetic equations. Simulations have been performed for head-on
$^{129}$Xe on $^{119}$Sn collisions at 32 AMeV.  The ingredients of 
simulations can be found in~\cite{I29-Fra01} as well as a detailed 
comparison between filtered simulated events (to account for the 
experimental device) and experimental data. 
A good agreement between both is revealed.

To refine the comparison higher-order charge correlations have been  calculated
for the simulated events, keeping the compact presentation  proposed
in~\cite{I31-Bor01}:  charge correlation functions are built for  all events,
whatever their multiplicity, by summing the correlated yields  for all $M$ and
by replacing the variable $\Zmoy$ by  $\Zbound = M \times \Zmoy = \sum_Z Z
n_Z$. Uncorrelated events are  constructed and weighted in proportion to real
events of each multiplicity. This presentation is based on the experimental
observation that the peaks 
observed independently for each fragment multiplicity correspond to the same 
$\Zbound$ region~\cite{I31-Bor01}. The variance bin was chosen equal
to one charge unit. We recall that in the considered domain of excitation
energy, around 3 MeV per nucleon~\cite{I11-Mar98,I29-Fra01}, secondary
evaporation leads to fragments one charge unit smaller, on average, than the
primary  $Z \approx 10 - 20$ ones, with a standard deviation around
one~\cite{T27Tab00}. If a weak enhanced production of exactly equal-sized
fragments exists, peaks are expected to appear in the interval $\sigma_Z =
0-1$, because of secondary evaporation. This interval in $\sigma_Z$ is hence
the {\it minimum} value which must be chosen to look for nearly equal-sized
fragments. Any (unknown) intrinsic spread in the fragment size coming from the
break-up process itself may enlarge the $\sigma_Z$ interval of interest. This
will be discussed at the end of the section, for the moment we will consider
only events with $\sigma_Z < 1$, which corresponds to differences of at most 
two units between the fragment atomic numbers in one event. 

Fig.~\ref{fig3} shows the correlation function calculated using the analytical
denominator (a) or the denominator given by the IPM (b).  Both functions are
drawn versus the variables $\Zbound= M \times \Zmoy$  and $\sigma_Z$. In
fig.~\ref{fig3}a, the equal-sized fragment correlations in the first bin are
superimposed over trivial correlations due to the finite size of the system.
For this reason, the ratio~(\ref{equ3}) is generally different from one and
smoothly varies with the variables $\Zbound$ and $\sigma_Z$. For each bin in
$\Zbound$ (fixed at 6 atomic number units), an exponential evolution of the
correlation function is observed from $\sigma_Z = 7 - 8$ down to $\sigma_Z = 2
- 3$. This exponential evolution is thus taken as a ``background'' 
empirically extrapolated down to the first bin $\sigma_Z = 0 - 1$
The amplitude of the
correlation function in the domain $\Zbound = 36 - 60$ is well above 
the background, with a confidence level higher than 90$\%$,  proving thus a
statistically significant enhancement of equal-sized fragment partitions. Of
the 1\% of events having $\sigma_Z < 1$, (0.13$\pm$0.02)\% 
(called extra-events from now on)
are in excess of the
background. In fig.~\ref{fig3}b, as expected, all correlations due to the
charge conservation are suppressed and the correlation function is equal to 1
(within statistical fluctuations)  wherever no additional correlation is
present. Again one observes peaks for $\sigma_Z < 1$. 
The percentage of
extra-events is 0.36$\pm$0.03\%, higher than the one obtained with the previous
method.
 Moreover, with this
method, peaks also appear at the maximum values of $\sigma_Z$ for a given
$\Zbound$. They correspond to events composed of one big (a heavy residue) and
several lighter fragments (sequentially emitted from the big one). In that case
fusion-multifragmention does not occur and the peaks reveal the small 
proportion 
(0.15\%)
of events which undergo the fusion-evaporation process. 
The IPM approach may thus reveal other correlations not seen with the
previous one.
Therefore, we shall use this approach in the following  analysis of the 
experimental data.     
\par
To conclude this part we can say that, although all events in the 
simulation arise from spinodal decomposition, only a very small fraction 
of the final partitions  have nearly equal-sized fragments. Different 
effects: beating of different modes, coalescence of nascent fragments, 
secondary decay of the excited fragments and, above all, finite size effects 
are responsible for this fact~\cite{Jac96,Colo97}. The signature of spinodal 
decomposition can only  reveal itself as a ``fossil'' signal.

\subsection{Experimental results\label{subsect43}}
\begin{figure}[htb]
\centering
\includegraphics[scale=0.8]{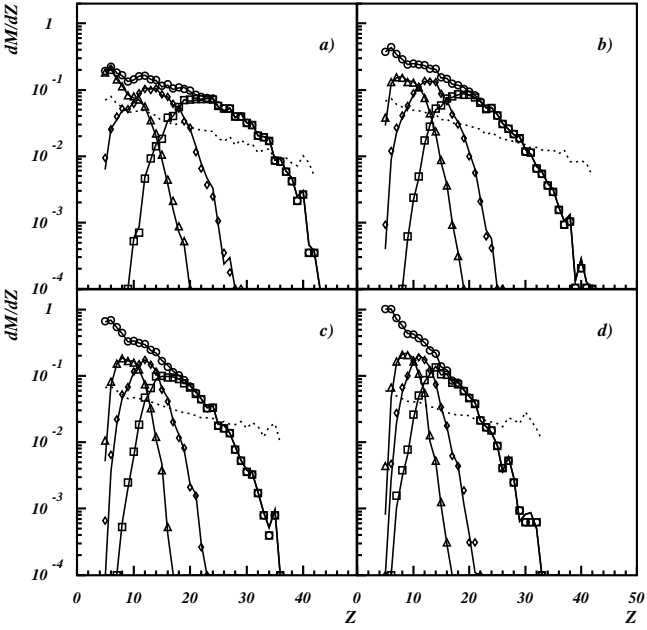}
\caption{Experimental differential charge multiplicity distributions
(circles) for the single source formed in central 39 AMeV
$^{129}$Xe on $^{nat}$Sn collisions. Parts a, b, c and d refer respectively to
fragment multiplicities 3, 4, 5, 6. The $Z$ distributions for the first 
(squares), second (diamonds) and third (triangles) heaviest fragments are 
presented too. The lines correspond to the results obtained with IPM. 
The dashed lines display the intrinsic probabilities.}
\label{fig4}
\end{figure}
We shall present now  higher-order charge correlations for the selected 
experimental events. This will be done for four incident energies 
(32, 39, 45 and 50 AMeV) in the framework of the IPM for the denominator.
The first step consists in determining the intrinsic probabilities of
fragments for each multiplicity and at each incident energy.  These 
probabilities are obtained by a recursive procedure of minimization. 
The minimization criterion is the normalized $\chi^2$ between the
experimental fragment probabilities and the fragment partition 
probabilities given by~(\ref{Eq P(N)}). The calculated $\chi^2$ were always 
lower than one: the lower is the incident energy and the larger is the 
multiplicity, the lower is the $\chi^2$ value.

\begin{figure}[htb]
\centering
\includegraphics*[scale=0.8]{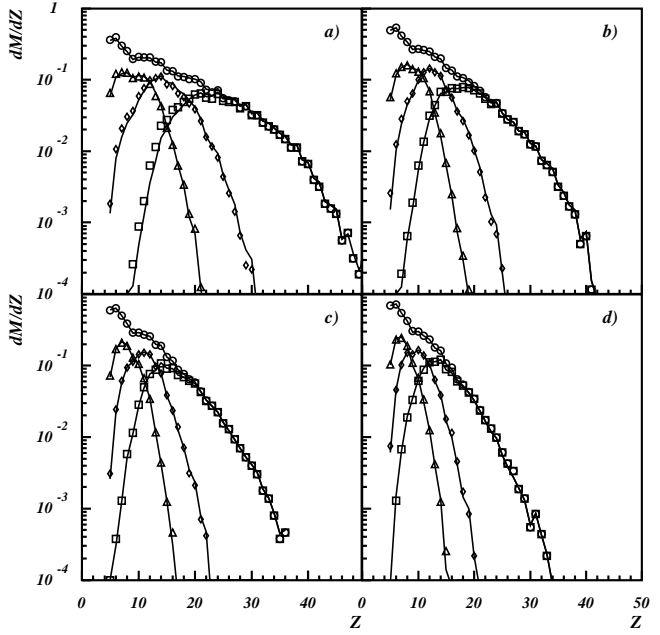}
\caption{Experimental differential charge multiplicity distributions
(circles) for the single sources formed in central 32, 39, 45 and 50 AMeV
 $^{129}$Xe on $^{nat}$Sn collisions (a, b, c, d). The $Z$ distributions 
for the first (squares), second (diamonds) and third (triangles) heaviest 
fragments are presented too. The lines correspond to the results obtained  
with IPM.}
\label{fig5}
\end{figure}
\begin{figure*}[htb] 
\centering
\includegraphics*[scale=0.6]{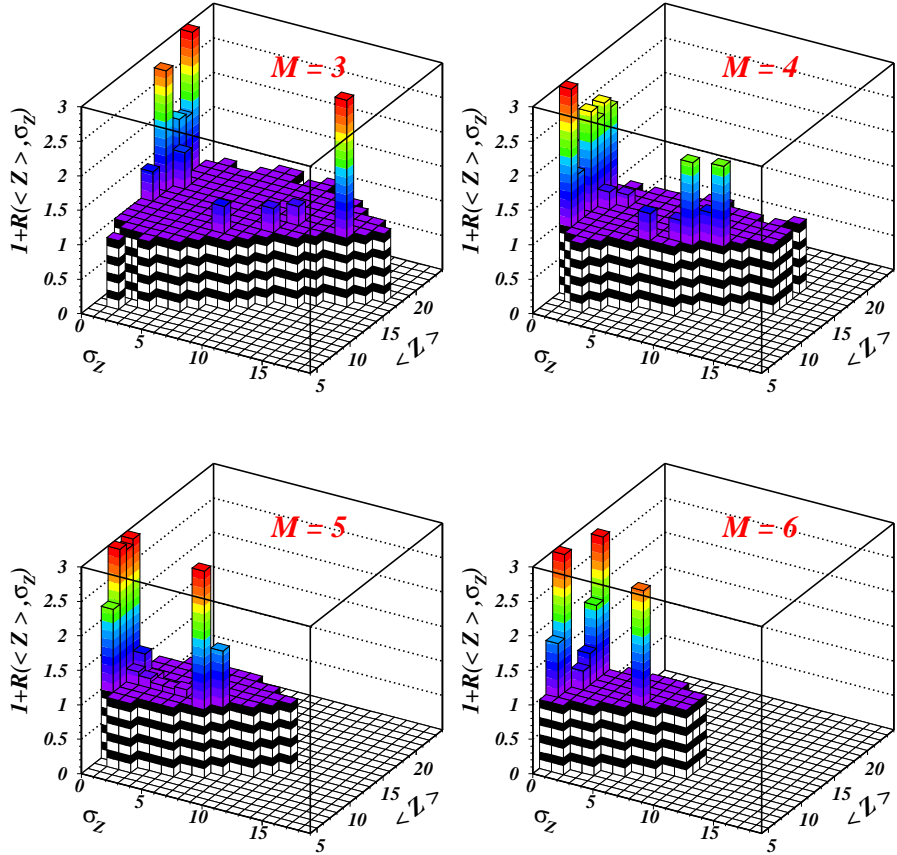}
\caption{Experimental higher-order charge correlations for selected events
formed in central 39 AMeV $^{129}$Xe on $^{nat}$Sn collisions, for
fragment multiplicities 3 to 6. The maximum value of the scale of the 
correlation function is limited to 3 on the picture.}
\label{fig6}
\end{figure*}
Charge distributions experimentally observed for the different fragment
multiplicities are shown  as an example in fig.~\ref{fig4}; they  correspond
to the 39 AMeV incident energy. Dashed lines refer to the  intrinsic
probabilities calculated with IPM and the corresponding charge distributions
are the full lines. The experimental charge distributions  are faithfully
described. Note that the ratio of the experimental to the  calculated curves
does not reveal any anomalous enhancement in a preferential  domain of fragment
atomic number. The charge distributions, summed over fragment multiplicities 3
to 6, are displayed in fig.~\ref{fig5}, for  the four incident energies. For
each incident energy the intrinsic  probabilities have been calculated
independently for the different fragment  multiplicities (see fig.~\ref{fig4})
and weighted in proportion to real  events of each fragment
multiplicity. The slight differences for the intrinsic probabilities
corresponding to the different multiplicities reveal small differences in the
average excitation energies of the multifragmenting sources (on the other hand,
it has been possible to fit all events from the BOB simulation, irrespective of
their multiplicity, using the same intrinsic probability distribution due to
the fact that the events correspond to the same initial conditions). We note
again the excellent agreement  between calculations and data at all incident
energies. The contributions to the Z distribution of the three heaviest
fragments of each partition are as well described as in fig.~\ref{fig4}, and
the  charges bound in fragments (not shown) are also perfectly reproduced.

Fig.~\ref{fig6} illustrates the higher-order correlation functions measured
for the different fragment multiplicities. It concerns single sources
selected at 39 AMeV incident energy. To make the effects more visible, peaks 
with confidence level lower than 80\% were flattened out.
We observe significant peaks in the bin $\sigma = 0 - 1$ for each fragment
multiplicity. For M=6, peaks are essentially located in the bin
$\sigma_Z = 1 - 2$. As observed in simulations, peaks corresponding to 
events composed of a heavy residue and light fragments ($\sigma_Z$ in the 
region 5-10 associated with low $\Zmoy$) are also visible.
\begin{table}
\caption{Characteristics of events with $\sigma_{Z} < 1$ for the different
incident energies. For each fragment multiplicity M, the range of $\Zmoy$ 
contributing to the correlation peaks are indicated (for bold $\Zmoy$ ranges, 
see text). The first line refers to BOB simulations (see
subsect.~\ref{subsect42}).}
\begin{center}
\begin{tabular}{lcccc}
\hline
   $M$ & 3 & 4 & 5 & 6 \\
     E (AMeV)  & & & &  \\
     \hline
32$*$& \textbf{12 - 20} & 9 - 17 & 8 - 13 & - \\
32   & \textbf{13 - 21} & 11 - 16 & -  & - \\
39   & \textbf{15 - 20} & 8 - 15 & 7 - 11 & 6 - 8 \\
45   & \textbf{17 - 18} & 5 - 14 & 8 - 11 & 6 - 8 \\
50   & - & - & - & 7 - 9 \\
\hline
\end{tabular} \\
\end{center}
\label{table2}
\end{table}
\begin{figure*}[htb] 
\centering
\includegraphics*[scale=0.6]{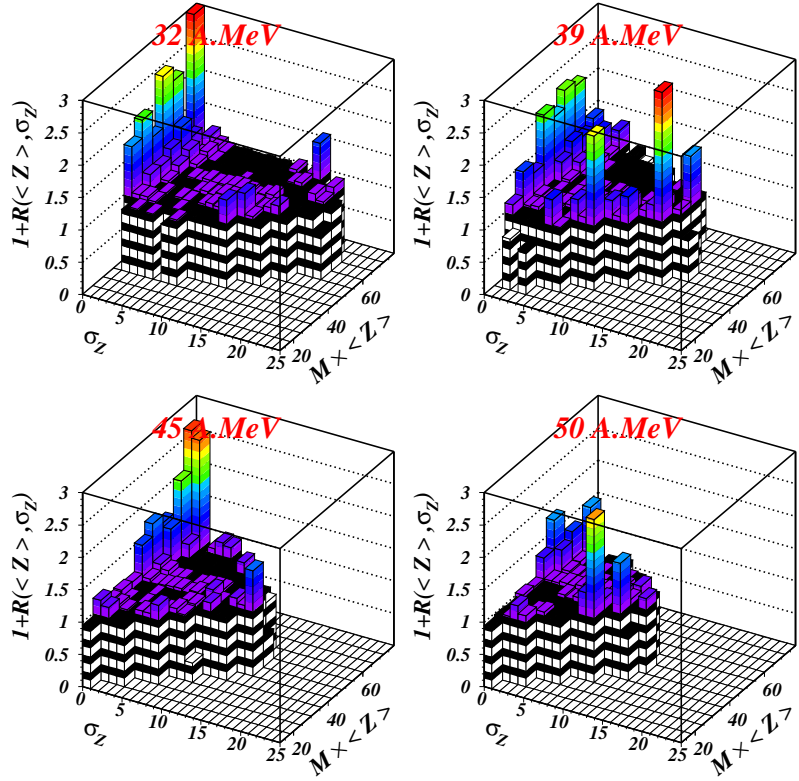}
\caption{Experimental higher-order charge correlations  for selected events
formed in central $^{129}$Xe  and $^{nat}$Sn collisions. Events with $M_f$=3
to 6 are mixed.}
\label{fig7}
\end{figure*}
\begin{figure*}[htb] 
\centering
\includegraphics*[scale=0.6]{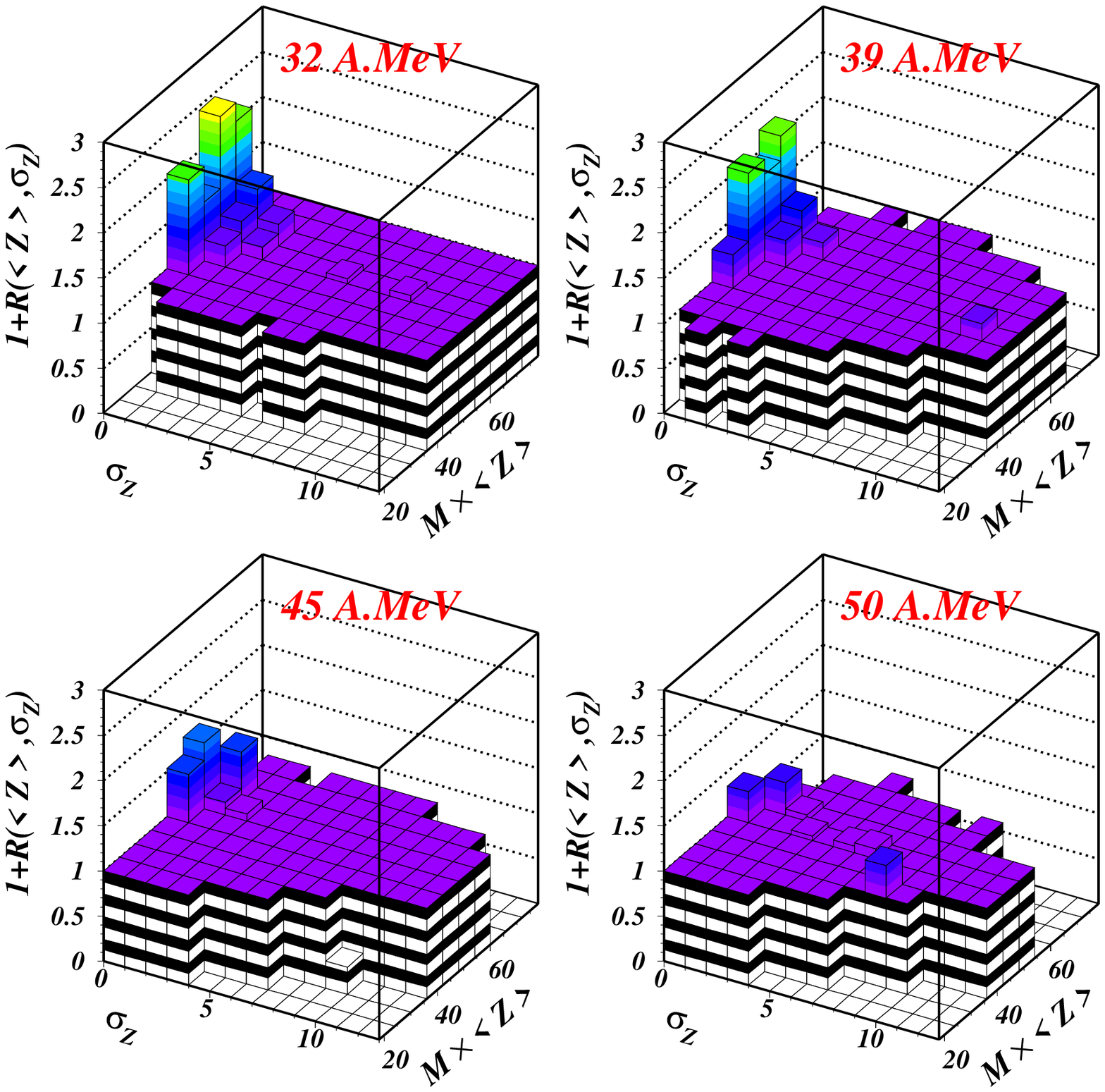}
\caption{Same as fig.~\ref{fig7} but peaks (or holes) with a confidence 
level lower than 90\% (1.65 $\sigma$) have been flattened out.}
\label{fig7b}
\end{figure*}

Let us now present the results for the different incident energies, summed over
multiplicities 3-6. We recall again that IPM denominators of correlation 
functions are weighted in proportion to real events of each multiplicity.  In
fig.~\ref{fig7} the measured functions  are displayed. If we exclude some
peaks with low confidence levels (as done in fig.~\ref{fig7b}) correlation
functions are equal or close to one except at low $\sigma_Z$ values. A summary
of average-charge domains contributing to  the correlation peaks in the first
bin in $\sigma_Z$ is given in Table  \ref{table2} as a function of fragment
multiplicity. All multiplicities, associated to the largest  $\Zmoy$ ranges
(20-6 and 18-6) contribute to the peaks at 39 and 45 AMeV. At 32 AMeV the peaks
only come from the low multiplicites ($M$ = 3-4), with a smaller $\Zmoy$ range
: 21-11;  at 50 AMeV, the situation is completely  different and we only
observe a contribution from $M$ = 6 with $\Zmoy$ = 7-9. Except in this latter
case, the $\Zmoy$ domains for which correlation peaks  are present are similar 
but slightly shifted  towards lower values with increasing energy.

\begin{figure}[htb]
\includegraphics[scale=0.7]{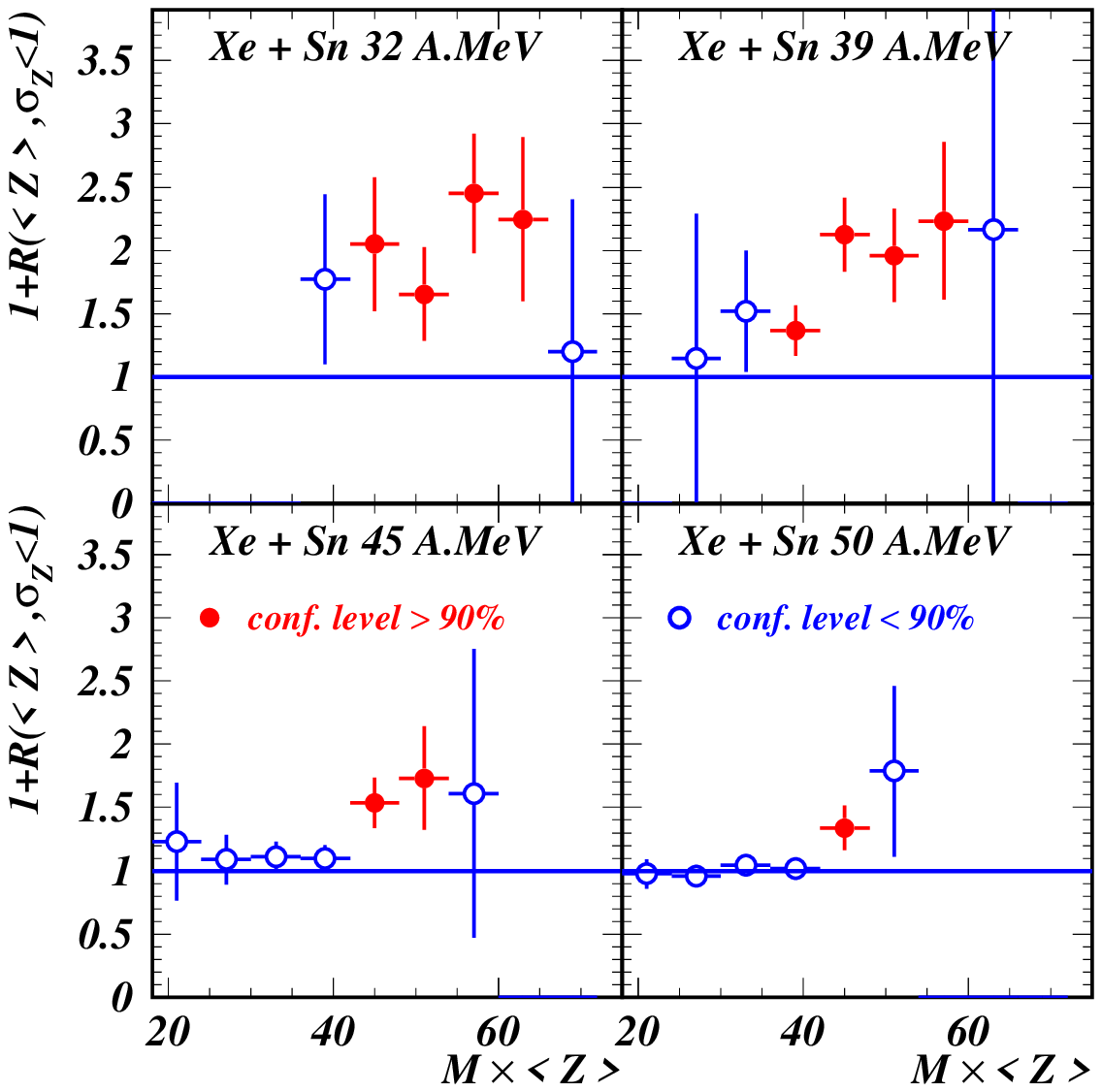}
\caption{Higher-order charge correlations at the different incident energies:
quantitative results for events with $\sigma_Z < 1$. Errors are
statistical assuming independent measurements and horizontal bars correspond 
to the M x $\Zmoy$ bins.}
\label{fig7t}
\end{figure}
To be quantitative, correlation functions for the first bin in $\sigma_{Z}$ are
displayed in fig.~\ref{fig7t} with their statistical errors. Full points
(open points) correspond to a confidence level higher (lower) than 90\%.
Two percentages of events relatively to the single source events
($M_f= 3-6$) are presented in columns 2 and 3 of Table~\ref{table3}: they
refer to the total number of events and of extra-events (taking into account
correlations and anti-correlations) in the bin. The present analysis fully
confirms at 32 AMeV incident energy the previous one~\cite{I31-Bor01} and
the extra-percentage of events with nearly equal-sized fragments is maximum
at 39 AMeV. 

\begin{table*}
\caption{
 Numbers and percentages of events  for $\sigma_Z < 1$ and
$\sigma_Z < 3$ and percentages of extra-events. The percentages are
expressed with respect to the total number of selected events with fragment
multiplicities 3 to 6. The first line refers to simulations (see
subsect.~\ref{subsect42}).}
\begin{center}
\begin{tabular}{lrccrcc}
\hline
  & \multicolumn{3}{c}{$\sigma_Z < 1$} & \multicolumn{3}{c}{$\sigma_Z < 3$} \\
  E   & \multicolumn{2}{c}{events}  & extra-events
&\multicolumn{2}{c}{events}  & extra-events  \\
 (AMeV)  &  & (\%) &  (\%) & & (\%) &  (\%) \\
\hline
32$*$& 353& 1.0  & 0.36 $\pm$ 0.03 & 4873& 14.0 & 1.30 $\pm$ 0.06 \\
32 &    83& 0.26 & 0.13 $\pm$ 0.02 & 1746&  5.5  & 0.79 $\pm$ 0.05 \\
39 &   151& 0.58 & 0.25 $\pm$ 0.03 & 4275& 16.3 & 1.29 $\pm$ 0.07\\
45 &   317& 1.32 & 0.21 $\pm$ 0.03 & 7328& 30.6 & 0.79 $\pm$ 0.05\\
50 &   762& 2.78 & 0.08 $\pm$ 0.02 & 12306& 45   & 0 \\
\hline
\end{tabular} \\
\end{center}
\label{table3}
\end{table*}
A closer examination of fig.~\ref{fig3}b and~\ref{fig7b} reveals an islet
of peaks, with a high confidence level, in the second and the third bins in 
$\sigma_Z$. They are located in the upper region of the  $M \times \Zmoy$ ranges
indicated in table~\ref{table2}. 
They correspond to events with a broader spread of charges. Observing such
events in the simulation may indicate that the intrinsic spread (which is
unknown) due to spinodal decomposition is larger than that coming from 
secondary evaporation, and hence that these peaks also sign the original 
process. The percentages of events with  $\sigma_Z < 3$ are also reported in
table~\ref{table3}. The conclusions are the same as above: while more
events have small values of $\sigma_Z$ when the incident energy increases,
the percentage of extra-events shows a maximum at 39 AMeV, and 
vanishes at 50 AMeV. 

\par\section{Discussion\label{sect5}}

The first hint of a bulk effect
for producing  fragments in central collisions between heavy nuclei at a given
moderate excitation energy ($\epsilon^* \approx 7 A$MeV) was the measurement
of identical fragment charge distributions for two different
system sizes (32 AMeV $^{129}$Xe + $^{nat}$Sn and 36 AMeV 
$^{155}$Gd + $^{nat}$ U)~\cite {I12-Riv98}. However this feature could
as well be interpreted from a statistical point of view, namely 
the dominance of phase space~\cite{Mor97,Bon95}. Indeed both dynamical
and statistical approaches were able to reproduce the experimental observation.
With the same dynamical approach as that used in this paper, the experimental
charge and multiplicities distributions were reproduced while the
average fragment kinetic energies were accounted for within 
20 \%~\cite{I29-Fra01}.
These same
properties were also well accounted for with the statistical model
SMM~\cite{Bon95,T16Sal97,LenBol00,TabBol00}.  From this agreement between data 
and the two models we learnt that the dynamics involved is sufficiently 
chaotic to finally explore enough of the phase space and describe fragment 
production through a statistical approach. To go further more constrained
observables were needed. Such was the goal of the present studies.

\begin{figure*}[htb]
\centering
\includegraphics[scale=0.8]{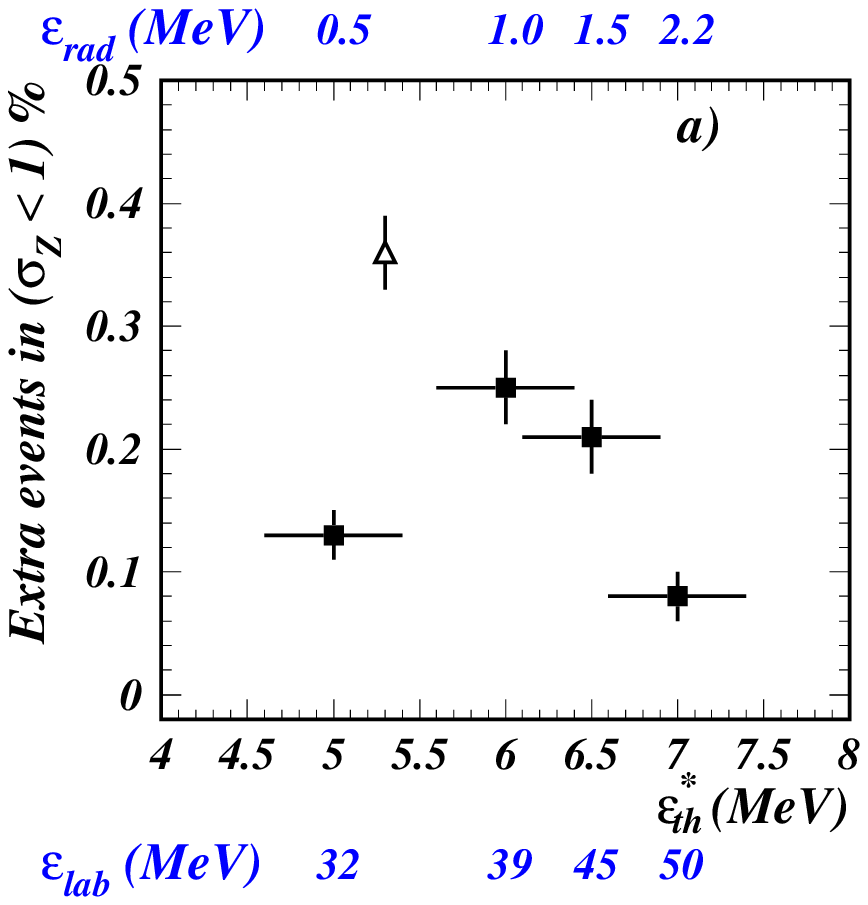}
\includegraphics[scale=0.8]{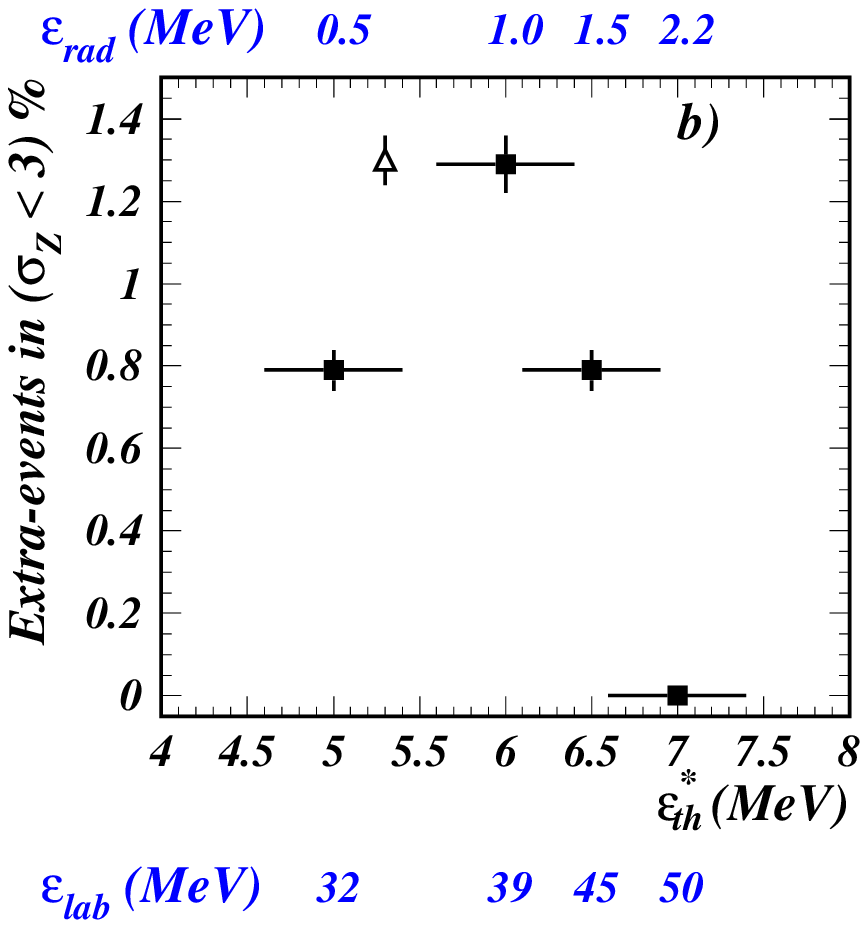}
\caption{Abnormal production of events with nearly equal-sized fragments
(a: $\sigma_{Z} < 1$ and b: $\sigma_{Z} < 3$) as a function of thermal
excitation energy (deduced from SMM): full points. The incident and radial
energy scales are also indicated.
The open point refers to the result from BOB simulations; the average thermal
excitation energy is used. Vertical bars correspond to statistical errors and
horizontal bars refer to estimated uncertainties on the backtraced
quantity, $\epsilon ^*$.}
\label{fig8}
\end{figure*}

 We have just seen that, at 32 AMeV incident energy, experimental correlation 
functions are similar to those obtained with events from dynamical
simulations, BOB. Above all, in both of them, there is abnormal production of 
nearly equal-sized fragments pointed out by the peaks in the first 
$\sigma_Z$ bin(s). Supported by this simulation we thus attributed the 
greatest part of fusion-multifragmentation events to spinodal 
decomposition~\cite{I31-Bor01}. These peaks hence appear as fossil 
fingerprints of the break-up process. 

Correlation peaks are also observed for experimental data at higher incident 
energies: the excitation function is displayed in fig.~\ref{fig8}.
Information on the associated thermal excitation energies (and extra radial 
collective energy) involved over the incident energy domain studied can be 
provided by the SMM model which well describes static and dynamics 
observables of fragments. Starting from a freeze-out volume fixed at three 
times the normal volume, the thermal excitation energies of the dilute 
and homogeneous system, extracted from SMM, vary from 5.0 to 7.0 AMeV and 
the added radial expansion energy remains low: from 0.5 to 2.2
AMeV~\cite{T16Sal97,T25NLN99}.
A rise and fall of the percentage of ``fossil partitions'' from spinodal
decomposition is measured. Fig.~\ref{fig8} reveals some difference 
between the experimental (full symbols) and simulated events (open
symbols): the experimental percentages of extra events are closer to the
simulated ones in fig.~\ref{fig8}b than in fig.~\ref{fig8}a.
This means that the charge distributions inside an event are slightly 
narrower in the simulation than in the experiment 
either because of the primary intrinsic spread, or because the
width due to evaporation is underestimated.
For the considered system, incident energies around 35-40 AMeV appear as 
the most favourable to induce spinodal decomposition; it corresponds 
to about 5.5-6 AMeV thermal excitation energy 
associated to a very gentle expansion energy around 0.5-1 AMeV. The 
qualitative explanation for those numbers can be well understood in terms of 
a necessary compromise between two times.
On one hand the fused systems have to stay in the spinodal region $\approx$ 
100-150 fm/c~\cite{Colo97,Idi94,Nor02}, to allow an important amplification 
of the initial fluctuations and thus permit spinodal decomposition; 
this requires a not too high incident energy, high enough however
for multifragmentation to occur. On the other hand, for a
finite system, Coulomb interaction and collective expansion push 
the ``primitive'' fragments apart and reduce the time of their mutual 
interation, which is efficient to keep a memory of ``primitive'' size 
properties. Above 45 AMeV incident energy part
of trajectories followed by the system in the temperature-density plane may
not sufficiently penetrate the spinodal region. Finally, considering secondary
decay effects, they
should be essentially similar between 39 and 50 AMeV 
because the primary fragment excitation energy per nucleon was found
constant over this incident energy range~\cite{I39-Hud02}. Particularly the
excitation function shown in fig.~\ref{fig8} should not be affected by such
effects.

Let us come now to the size of fragments associated to ``fossil partitions''.
From the theoretical point of view, spinodal instabilities have been mainly
studied within semiclassical (as BOB) or 
hydrodynamical~\cite{Nor02,Nor00,Hei88} frameworks. Precise information on 
the most unstable collective modes, which is needed for discussing fragment 
size produced by spinodal decomposition, can only be obtained when
considering quantal effects. From the limited number of studies investigating 
such effects~\cite{Ayi95,ChoHirs99,Col02} a few trends emerge: i) an
increase of the number of unstable collective modes with the size of
the system (up to multipolarity $L$=5-6 for A=140), ii) a decrease of the
$L_{max}$ value when the temperature increases and iii) a dominance
of the octupole mode ($L$=3) which appears as the most unstable.
Charge correlations were only studied for events with 3 to 6 fragments,
which precludes any information on the quadrupole mode ($L$=2).
If we except results at the highest incident energy, which strongly differ
from others, we do observe the dominance of events with three fragments,
associated to the multipolarity $L$=3. Indeed the $Z$ domains
marked in bold in table~\ref{table2} correspond to the largest proportion of
events which populate the first $\sigma_Z$ bin, which vary from 70\% at 32
AMeV to 40\% at 45 AMeV. Then the number of events with higher multiplicities,
corresponding to larger $L$ values, progressively decreases, as expected.

 At 50 AMeV results observed for the fragment sizes ($\sigma_Z < 1$) are
 puzzling. Moreover 
the percentage of
extra-events drops to zero when events with  $\sigma_Z < 3$ are considered.
Further simulations and theoretical works are needed to progress in the 
interpretation of these data.

\section{Conclusions\label{sect6}}
In conclusion, we have investigated charge correlation functions for compact
and heavy fused systems which undergo multifragmentation, as a function of
the incident energy, from 32 to 50 AMeV. 
At the lowest energy, we have confirmed an enhanced production of events with 
equal-sized fragments. Supported by theoretical simulations we have 
interpreted this enhancement as a signature of spinodal instabilities as the 
origin of multifragmentation of those systems in the Fermi energy domain. 
This fossil signal culminates for incident energy around 35-40 AMeV, which 
corresponds to the formation of a hot and dilute system at 0.3-0.4 the 
normal density and temperature around 4-5 MeV~\cite{TGua96,I29-Fra01}.
Spinodal decomposition describes the dynamics of a first order phase 
transition, and the present observations support the existence of such a
transition for hot finite nuclear matter~\cite{MDA00,LenBol00,MDA02}.

\begin{acknowledgement}

The experimental results were obtained  by the INDRA collaboration.
\end{acknowledgement}

\section*{Appendix}

\begin{figure}[htb]
\includegraphics[scale=0.7]{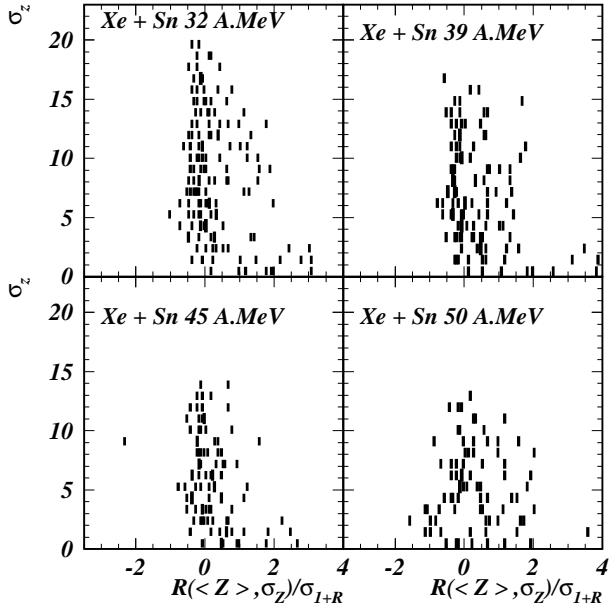}
\caption{Deviations from 1 of the correlation functions divided by the
statistical errors in abcissa, for the different values of $\sigma_Z$.
Correlation functions calculated by the IPM method, and shown in
fig~\ref{fig7}
}
\label{fig11}
\end{figure}
\begin{figure}[htb]
\includegraphics[scale=0.7]{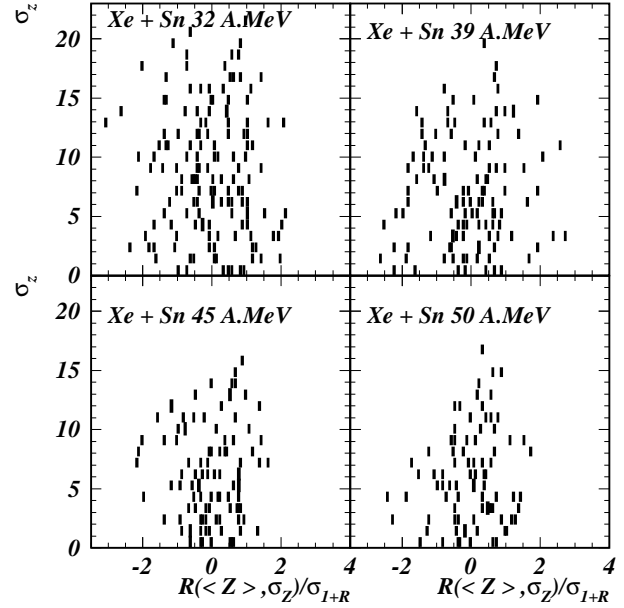}
\caption{Deviations from 1 of the correlation functions divided by the
statistical errors in abcissa, for the different values of $\sigma_Z$.
Correlation functions calculated by the exchange mixing event method.
}
\label{fig12}
\end{figure}
Very recently a new method for building charge correlation functions
was proposed, of implementation easier than the IPM
method: the denominator is built by mixing events through random
exchanges of two fragments between two events under the constraint that
the sum of the two exchanged fragments is conserved, besides the
conservation of $\Zbound$~\cite{I41-Cha02}.
To test the sensitivity ot this method as compared to the IPM method, the
present experimental data were also analyzed using the exchange method, by
building correlation functions $1 + R(\sigma_Z, \Zmoy)$ with $\sigma_Z$ being
calculated from eq.~\ref{equ2}.
The results of the two methods are compared in fig.~\ref{fig11} and
fig.~\ref{fig12}, displaying
for each bin of the plane ($\sigma_Z$, M $\times \Zmoy$), the deviation from 1,
$R(\sigma_Z, \Zmoy)$, of the correlation function, normalized to its
statistical error bar, $\sigma_{1 + R(\sigma_Z, \Zmoy)}$, calculated from
the numerator only.

 The greater sensitivity of the IPM method is clearly seen on 
 fig.~\ref{fig11}: the values of the
correlation function are closer to one ($R$=0) except at low $\sigma_Z$
where we observe correlations with a significant confidence level. Conversely,
the exchange method, fig.~\ref{fig12}, leads to a large dispersion of the 
values of $R(\sigma_Z, \Zmoy)/\sigma_{1 + R(\sigma_Z, \Zmoy)}$, $\sim$1.6  
times broader than with the IPM method at 32, 39 and 45 AMeV. At 50 AMeV
both methods lead to similar dispersions.

%
%

\end{document}